\documentclass[3p, times, twocolumn]{elsarticle}

\usepackage{amssymb}

\usepackage[figuresright]{rotating}

\begin{document}
\bibliographystyle{elsarticle-num}

\begin{frontmatter}





\title{Theory @ Hard Probes 2015}

\author{Bj\"orn Schenke}

\address{Physics Department, Brookhaven National Laboratory, Upton, NY 11973, USA}

\begin{abstract}
Overview of the latest theory developments presented at the Hard Probes 2015 conference, held at McGill University, Montreal, Canada, in July 2015.
\end{abstract}


\end{frontmatter}


\section{Introduction}
\label{seq:intro}
The study of hard probes in heavy ion collision is equivalent to any standard experiment where a calibrated probe is used, i.e., a
physical object under strict theoretical control for which a definite relationship between its properties and those of the probed system can be established. Unfortunately, we cannot hit the matter produced in heavy ion collisions, namely the quark gluon plasma, with an external probe. It lives for a too short period of time and its extent typically does not exceed $\sim 20{\rm fm}$. Thus, one uses the hard probes that are produced in the heavy ion collision itself. This includes high momentum hadrons as well as fully reconstructed jets. Also electromagnetic probes, like photons and dileptons, of various energies and momenta should be useful tools to learn about the early stages of the collision.

The goal of these measurements is to learn about the strong force, namely the theory of quantum chromo dynamics (QCD), and the properties of extended hot and dense systems governed by it. In particular we seek to experimentally access the phase diagram of QCD, which contains information on where and how deconfinement and chiral symmetry restoration occurs. Further, we aim at gaining insight into the transport properties of the QCD medium produced, in particular characterized with regards to hard probes by $\hat{q}$, the average transverse momentum squared per unit length acquired by a hard particle traversing the medium. This is related to other transport properties of the medium, most prominently the shear viscosity to entropy density ratio $\eta/s$ and the bulk viscosity to entropy density ratio $\zeta/s$, which in principle are also accessible via the study of soft particles and their correlations.

In this overview we review the latest progress in the field of heavy ion physics presented at the Hard Probes 2015 conference. Because the study of medium effects on hard probes involves not only the calculation of the hard probes and their evolution, but also requires a firm understanding of the initial state and bulk evolution of the medium, both of these subjects were covered at the conference. We further discuss heavy flavor and electromagnetic probes, as they provide additional insight into the medium properties, at any momentum scale. Since recently, the study of high multiplicity events in small collision systems such as p+p and p+A has shown that they contain interesting physics beyond expected cold nuclear effects. We thus devote a section to these systems and discuss our latest understanding of the measurements.

We organize this overview by sub-field, beginning with a description of the current status of describing the initial state of heavy ion collisions, followed by new developments in the sector of bulk evolution and electromagnetic probes. We then move on to discuss the latest results on hard probes themselves, and present an overview of heavy flavor physics relevant for heavy ion collisions. Finally we will present the theoretical status of the description of high multiplicity events in small collision systems.

\section{Initial state}
\label{seq:initial}
The initial state description of heavy ion collisions, namely the calculation of the coordinate and momentum space structure of the produced matter, has advanced quite significantly in the past years. In particular the calculation of the low momentum region, which is most relevant in determining the initial energy momentum tensor of the hydrodynamic evolution, is challenging and only recently has there been progress in obtaining it from a Glasma initial state, described by classical gluon fields within an effective theory of QCD \cite{Schenke:2012wb,Schenke:2012hg,Gale:2012rq}. At this conference, an initial state description using a somewhat different philosophy, starting with pQCD and introducing saturation to regulate the low momentum part, was presented. This EKRT (after the authors of \cite{Eskola:1999fc}) framework has several features in common with the IP-Glasma and is the only other model shown to describe the complete event-by-event distributions of the flow harmonics \cite{Niemi:2015qia}. Since it lacks the fine structure in the transverse spatial energy density distribution, this indicates that such details are less important than the larger scale prescription for energy deposition to describe the observables studied so far at RHIC and LHC.
We will return to this model in the next section when we discuss the bulk observables.

Another important aspect of the initial state is the description of the incoming nuclei and their gluon distribution. Recent progress in this direction is the solution of the next to leading order (NLO) BK equation \cite{Balitsky:2008zza}. It turns out that numerical solutions lead to unphysical results, namely solutions that are not positive definite, depending on the initial condition \cite{Lappi:2015fma}. This behavior is seemingly cured when resumming large radiative corrections that are enhanced by double collinear logarithms \cite{Iancu:2015vea}. The general result is a slowing down of the evolution due to NLO corrections.

\section{Bulk properties}
\label{seq:bulk}
In addition to the initial state physics, the theoretical description of the medium evolution is a major component in modeling heavy ion collisions and necessary to determine the transport properties of the medium in comparison with experimental data. It is further a necessary ingredient in all calculations of hard probes, if one wants to compare to experimental data. 

Viscous relativistic fluid dynamics is the most successful tool in modeling the bulk evolution in heavy ion collisions \cite{Gale:2013da}. 
Extended theories of relativistic fluid dynamics, which are necessary because relativistic Navier-Stokes theory is unstable due to acausally propagating modes, can be derived from the relativistic Boltzmann equation. Relatively new derivations (DNMR after the authors), that improve upon the traditional 14 moment approximation, provide a more physical truncation by only considering the relevant (slowest) time scales and using a systematic power counting in Knudsen and inverse Reynolds numbers \cite{Denicol:2012cn}. Further new developments include the so called anisotropic hydrodynamics, where the gradient expansion is performed around a modified (anisotropic) momentum distribution \cite{Martinez:2010sc,Martinez:2012tu}. This leading order ``aHydro'' was extended to also include corrections that are not governed by included anisotropy of the momentum distribution, and dubbed vaHydro \cite{Bazow:2013ifa}. A recent comparison shows that the latter leads to best agreement with an exact solution of the Boltzmann equation, followed by aHydro, DNMR and the conventional 14 moment method
\cite{Denicol:2014xca,Denicol:2014tha,Bazow:2015cha}.

As mentioned in the previous section, the EKRT framework, together with second order viscous hydrodynamics was used to describe a wide range of observables and constrain the temperature dependence of $\eta/s$ \cite{Niemi:2015qia,Eskola:2015uda}. Because of different sensitivities of various observables to the initial state and the transport parameters of the medium, a strategy for constraining both is emerging. The initial state and its fluctuations can be constrained by matching the event-by-event distributions of the flow harmonics (which are insensitive to the transport parameters). Then the mean values of the $v_n$ and the event plane correlations at different energies can be used to constrain $(\eta/s)(T)$ and potentially $(\zeta/s)(T)$.

There are of course alternative descriptions to hydrodynamics. More microscopic descriptions of the medium, e.g. numerical solvers of the relativistic Boltzmann equation, have also been used to determine transport parameters such as $\eta/s$ \cite{Plumari:2015cfa}. In this framework, a similar sensitivity of the flow harmonics to the temperature dependence of $\eta/s$ at different energies as determined in hydrodynamic calculations \cite{Niemi:2012ry,Niemi:2015qia} was found.
 
\section{Electromagnetic probes}
Recent focus in the sub-field of electromagnetic probes in heavy ion collisions has been on the understanding of the yields and anisotropic flow of photons. 
Many calculations, and in particular hydrodynamic models that describe hadronic observables extremely well, underestimate the experimental data on photon yields and flow from RHIC or LHC. 

Several suggestions to improve the agreement with data by including thus far ignored but potentially important physical effects have been made. One of them is the possible chemical non-equilibrium at early times, leading to delayed quark production and thus a delay of photon production to later times \cite{Monnai:2014kqa}. This increases the photon $v_2$ because photons will be produced at times when the system has already built up more flow. However, yields are further reduced, increasing the problem.

Improvements of the thermal photon calculation using hydrodynamic evolution, in particular the inclusion of bulk viscosity and more complete hadronic photon rates, have also led to reduced tension between the theoretical photon yield and $v_2$
and the experimental data \cite{Paquet:2015lta}. The later stage photon production was identified to be very important (see also \cite{Linnyk:2013wma,Linnyk:2015tha}), and further improvements in the description of hadronic photon production rates will be necessary. Likewise, a more sophisticated treatment of jet-medium photon production will be required.

Anisotropic hydrodynamic calculations have explored the effect of the initial momentum-space anisotropy of the QGP on photon and dilepton production \cite{Ryblewski:2015hea,Bhattacharya:2015ada,Bhattacharya:2015wca}. 
Photon elliptic flow is found to increase with increasing $\eta/s$ in this study, regardless of the assumed initial momentum-space anisotropy, whose variation leads to a modification of the shape of $v_2$ vs.~transverse momentum.

Another recent development is the calculation of thermal photon and dilepton rates to next to leading order (NLO) in \cite{Ghiglieri:2013gia,Ghiglieri:2014kma,Ghiglieri:2015nba}. Of particular interest is the generalization of the dilepton rate to arbitrary invariant mass $M$ by combining the results from \cite{Ghiglieri:2014kma} and \cite{Laine:2013vma}. An interesting conclusion from this work is that the rates are only mildly affected by NLO corrections, even for $\alpha_s = 0.3$, and for this coupling are smooth across the light cone, which supports the use of low-mass dileptons as a proxy for real photon measurements.

Dileptons can further be used to gain important insight into chiral symmetry restoration and whether it is achieved in heavy ion collisions. One clear indication of this would be a measurement of degenerate $\rho$ and $a_1$ meson spectral functions. Unfortunately, due to the broad structures in the axial channel and large backgrounds, the $a_1$ spectral function is very hard to measure.
So a direct demonstration of degenerate spectral functions of chiral partners is not possible at the moment. However, it has been shown \cite{Hohler:2013eba,Hohler:2015mda} that the broadening of the $\rho$ meson spectral function, which is calculated using Weinberg sum rules and necessary to describe the dilepton invariant mass spectra in heavy ion collisions, is compatible with chiral symmetry restoration. More precisely, the mechanism of chiral symmetry restoration is an evaporation of the chiral $\rho-a_1$ mass splitting with increasing temperature and increasing low energy strength (close to the $\rho$ mass) in the $a_1$ spectral function along with the broadening of the $\rho$.

\section{Heavy flavor}
\label{seq:heavy}
The medium modification of quarkonium production could give important information on the formation of the quark gluon plasma and its detailed properties. Therefore it is important to develop a sophisticated understanding of quarkonium production, one aspect of which is the consideration of formation time \cite{Song:2015bja}. The resulting delay of quarkonium production will increase its survival probability in a hot and dense medium, since it will be produced at a lower temperature where the thermal decay width is smaller. 
Recombination is another important aspect of quarkonium production in heavy ion collisions, and its presence is supported by measurements of the $R_{AA}^{J/\psi}$ at LHC \cite{Abelev:2012rv}. Furthermore, the production of excited charmonium states in comparison to the $J/\psi$ meson could give additional information on the recombination process. This was studied in \cite{Cho:2014xha} by evaluating the Wigner functions of the $J/\psi$ and $\psi(2S)$ states. Here it was argued that differences in the wave function distributions in momentum space of the two states can explain the measurement of the nuclear modification factor ratio between the $\psi(2S)$ and $J/\psi$ meson \cite{Khachatryan:2014bva}.

Considering open heavy flavor, the detailed description of the energy loss of the heavy quarks is important to describe e.g. the $R_{AA}$ and elliptic flow $v_2$ of D mesons. It was found recently that the variation of the temperature dependent drag coefficient (as well as the inclusion of coalescence) can make a big difference for the elliptic flow coefficient of $D$ mesons, while the $R_{AA}$ can be relatively unaffected \cite{Das:2015ana,Scardina:2015fxa}. This allows for a simultaneous description of the measured $R_{AA}$ and $v_2$ \cite{Adare:2006nq}.
In medium hadronization was also pointed out to be of importance to achieve this agreement within the POWLANG framework in \cite{Beraudo:2014boa,Nardi:2015pca}.

To study heavy quark energy loss, it was pointed out in \cite{Huang:2015mva,Xing:2015mua} that tagging jets with B-mesons is very helpful, because it enhances the sample of events with heavy quarks produced at the early stages of the collision significantly.
The quenching of such b-tagged jets was computed in above works using the GLV energy loss formalism and predictions are presented for $R_{AA}$ and the momentum imbalance. 

Regarding radiative energy loss of heavy quarks, within the Higher Twist framework the drag induced radiative energy loss of heavy quarks was found to be just as important as that induced by transverse momentum diffusion \cite{Abir:2015hta}.

Medium effects on heavy quarks can also be obtained from lattice calculations, which have yielded updated estimates of the heavy quark diffusion coefficient $DT = 0.59...1.1$ \cite{Francis:2015daa}. The lattice can further provide the medium modification of quarkonium states. Recent limits on the survival of bottomonium states from lattice NRQCD show survival of the $\Upsilon$ and $\chi_{b1}$ states up to $T\gtrsim 249\,{\rm MeV}$ \cite{Kim:2014iga}.

\section{Parton showers and jets}
\label{seq:hard}
To describe jet measurements at the LHC and RHIC, theoretical calculations need to go beyond the energy loss and momentum diffusion of single partons, and describe the entire parton shower. This involves describing the virtuality (or angular) ordered vacuum shower, the time ordered in-medium shower and possible interferences between the two.
The vacuum shower is generally under good control and can be simulated realistically using Monte Carlo techniques by event generators \cite{Sjostrand:2009ad} such as PYTHIA \cite{Sjostrand:2006za}, HERWIG \cite{Corcella:2000bw}, or SHERPA \cite{Gleisberg:2008ta}. 

While at high virtuality (and high energy), corrections to the vacuum shower can be accounted for in the Higher Twist (HT) formalism \cite{Guo:2000nz}, at low virtuality (and high energy) the correct framework is the BDMPS-Z description \cite{Baier:1996kr,Baier:1996sk,Zakharov:1996fv,Zakharov:1997uu}.
The transition from one regime to the other is however not easy to describe and more details on the different regimes and the issue of their combination can be found in a dedicated contribution in these proceedings \cite{Majumder:2015qqa}.
Further, at low energy, perturbative methods seize to be valid: the radiated soft partons (of order $\sim 1\,{\rm GeV}$) are strongly
coupled to the medium and the only available method to describe this regime is via the conjectured AdS/CFT correspondence \cite{Gubser:2006bz,Arnold:2010ir,Arnold:2011qi}. This method has been combined with a vacuum shower in \cite{Casalderrey-Solana:2014bpa} and comparison to data has shown good agreement \cite{Casalderrey-Solana:2015vaa} for various jet observables. 

Analyzing the time scales in the BDMPS-Z formalism, the structure of a medium shower in the high energy and low virtuality regime was shown to consist of a coherent inner core and large angle decoherent gluon cascades \cite{CasalderreySolana:2012ef,Blaizot:2015lma}.
Furthermore, color decoherence removes the angular ordering of the vacuum shower and allows for additional soft radiation within the inner cone \cite{Apolinario:2014csa}.
QGP-induced successive branchings along the parton shower are independent and quasi-local (interferences are suppressed), which allows for a probabilistic picture of the shower evolution, as implemented in event generators such as MARTINI \cite{Schenke:2009gb}. In the multi-branching sector, the inclusive gluon distribution obeys a rate equation with a scaling solution, that describes the constant flow of energy down to the medium temperature scale \cite{Blaizot:2015jea,Blaizot:2015lma}.

The soft radiated partons lose energy and are ``broadened'' until the eventually thermalize. To study the details of this process, the jet evolution and thermalization can be combined via drag and diffusion in the Fokker-Planck equation \cite{Iancu:2015uja}. A characteristic structure of the (partially) quenched jet with the leading particle front and a soft thermalized tail is found.

The medium property that affects the momentum broadening of high momentum partons is the mean transverse momentum squared per unit length $\hat{q}$. The $\langle p_T^2 \rangle$ from a radiative process with one single scattering
picks up two logarithmic contributions (in $\omega$ and $k_T$). This enhancement can be absorbed into a redefinition of the jet quenching parameter $\hat{q}$ \cite{Liou:2013qya,Blaizot:2014bha}.

Another recent effort is to compute parton energy loss at NLO. This involves $\mathcal{O}(g)$ corrections to the drag from non-linear interactions of soft gluon fields and corrections to the collision kernel and the inclusion of (semi-collinear) wider angle bremsstrahlung for the radiative part \cite{Ghiglieri:2015zma,Ghiglieri:2015ala}. Without a detailed, possibly Monte Carlo, calculation it is hard to tell how much impact the NLO corrections will have on any observables.

Regarding the jet-medium interaction, from phenomenology, it was found to depend on temperature and the energy of the jet - otherwise a simultaneous description of $R_{AA}$ and $v_2$ is not possible \cite{Betz:2014cza,Betz:2015fra}. Further, the recoil of the jet on the medium should have a noticeable effect on the intra-jet structure \cite{Tachibana:2014lja,He:2015pra}.
This structure can also be described within soft collinear effective theory when medium interactions are included via ``Glauber gluon'' interactions \cite{Chien:2015hda}.

Finally, some progress is being made in the description of the last step in jet shower calculations, the hadronization. Although very hard partons should hadronize outside the medium, the highly abundant soft components of a jet have a high probability of hadronizing in the medium. Including thermal partons from the background medium into the hadronization is thus necessary and leads to a modification of the hadron distribution functions \cite{Han:2012hp}.

\section{Small systems}
\label{seq:small}
The study of small collision systems has gained increased attention due to the discovery of ridge like correlations in p/d+A and even p+p collisions. These may originate from correlations in the initial particle production or final state collective effects (for a recent review see \cite{Dusling:2015gta}).  

In particular, it was shown that initial state effects are able to produce a $v_2$ component for gluons \cite{Lappi:2015vha}, while a finite $v_3$ can be generated by the Yang-Mills evolution of the gluon fields \cite{Schenke:2015aqa}. For more details on this topic see \cite{Lappi:2015jka} in these proceedings.

If final state effects are not the dominant mechanism, forward $J/\Psi$ production in p+A collisions can tell us about ``cold nuclear matter'' effects, that can either be quantified by nuclear parton distribution functions (pdf) or calculated from the color glass condensate effective theory (CGC). Very detailed analyses were presented using nuclear pdf's at leading order and NLO, where within the uncertainties both can describe the experimental data on the $R_{\rm pPb}$ \cite{Vogt:2015uba}. However, the forward-backward ratios, independent of the interpolated p+p baseline, are not as well described. The somewhat older EKS98 LO parametrization was found to do the best job of describing all the data.
A new CGC calculation \cite{Ducloue:2015gfa,Ducloue:2015xqa} using a running coupling BK evolved dipole operator with parameters constrained by HERA data and Glauber nuclear geometry leads to better agreement with the experimental data on $R_{\rm pPb}$ than previous results \cite{Fujii:2013gxa}.
Furthermore, it was presented that coherent energy loss (the regime where the formation time $t_f > L$, with $L$ the length of the medium) can also describe the $R_{\rm pPb}$ (and $R_{\rm dAu}$) of $J/\psi$ measured at LHC (RHIC) \cite{Arleo:2015lja}. 

Concerning the ridge effect, it was suggested \cite{Kovner:2015rna} that within a hybrid approximation, treating the projectile proton in the parton model and the target nucleus in the CGC approach, also di-photon correlations should show a ridge like correlation in p+A collisions at RHIC energies. This type of correlation results from the double fragmentation component of prompt di-photons. A measurement of these correlations could provide complementary information to understand the underlying mechanism that produces the ridge phenomenon.

Finally, the $R_{pA}$ of jets in p+A collisions shows some interesting centrality dependence \cite{ATLAS:2014cpa}, which can be explained by the bias introduced by the presence of a large $x$ parton in the projectile, which limits the amount of low $x$ partons to contribute to the underlying event multiplicity \cite{Alvioli:2014eda,Perepelitsa:2014yta,Armesto:2015kwa}. 

\section{Summary}
In summary, a lot of progress is being made in using hard probes to learn about the properties of QCD matter created in heavy ion collisions. We are closer to a clear understanding of the detailed interactions of hard partons with the quark gluon plasma and the properties of in-medium showers. A combination of all relevant aspects of the medium evolution, hard parton production and evolution, as well as electromagnetic probes and heavy quark dynamics into a consistent framework is still outstanding, but will be achievable in the future. In combination with the analysis and theoretical description of small collision systems, this will lead to a detailed understanding of QCD under extreme conditions and the fundamental properties of matter.

\section*{Acknowledgments}
The author thanks Gabriel Denicol for comments on the manuscript. The author is supported under DOE Contract No. DE-SC0012704 and acknowledges a DOE Office of Science Early Career Award.




\bibliography{spires}







\end{document}